\documentstyle[12pt,aps,prd]{revtex}
\begin{document}
\tightenlines 
\vspace{.5cm}
\begin{center}
{\large\bf Extended supersymmetry for the Bianchi-type cosmological models}
\vspace{.5cm}\\

J.J. Rosales ${}^{a,b}$\footnote{e-mail: juan@ifug3.ugto.mx}, V.I.
Tkach ${}^b$\footnote{e-mail: vladimir@ifug3.ugto.mx} and J. Torres
${}^b$ \footnote{e-mail: jtorres@ifug1.ugto.mx} 
\vspace{.5cm}\\  
${}^a$ {\it Facultad de Ingenier\'{\i}a Mec\'anica El\'ectrica y 
Electr\'onica, Univ. de Gto. \\
Av. Tampico 912, C.P. 36730. Apdo. Postal 215-A\\
Salamanca, Gto. M\'exico}\\

${}^b$ {\it Instituto de F\'{\i}sica,  Universidad de Guanajuato\\
 C.Postal 66318, 05315-970 Le\'on, Gto. M\'exico}

\vspace{.5cm}

{\bf Abstract}
\end{center}

In this paper we propose a superfield description for all 
Bianchi-type cosmological models. The action is invariant under 
world-line local $n=4$ supersymmetry with 
$SU{(2)}_{local}\otimes SU{(2)}_{global}$ internal symmetry. Due to the 
invariance of the action we obtain the constraints, which form a closed 
superalgebra of the $n=4$ supersymmetric quantum mechanics. This 
procedure provides the inclusion of supermatter in a sistematic way.


\section{Introduction}
In the absence of a fundamental understanding
of physics at very high energies and, in particular, in the
absence of a consistent quantum theory of gravity, there is no hope, at 
present, to meet an understanding of the quantum origin of the Universe in a 
definite way. However, in order to come nearer to this presently unattainable 
goal it appears desirable to develop highly simplified, but consistent models,
which contain as many as possible of those features which are believed to
be present in a future complete theory. Spatially homogeneous 
minisuperspace models obtained by dimensional reduction from (1 + 3) to
(1 + 0) dimensions have, therefore, played an important role in quantum 
cosmology.$^1$ On the other hand, there are several reasons for studying 
locally 
supersymmetric theories rather than non-supersymmetric ones. Four-dimensional 
model with local supersymmetry called supergravity (SUGRA) theory, 
leads to a constraint 
which can be thought of as square root of the Wheeler-DeWitt constraint, and 
it is related to it in the same way as the Dirac equation is related to the 
Klein-Gordon equation.$^2$ However, due to the technical complexities, 
the early papers on canonical supergravity $^3$ make no attempt at 
exploting this idea, but content themselves with setting up the canonical 
formalism and discussing the classical constraint algebra in terms of 
Poisson (or Dirac) brackets. 

In the case of SUGRA theories one can find one-dimensional
supersymmetric quantum mechanics (SQM) models by reducing four-dimensional 
$N=1$ SUGRA coupled to supermatter.$^4$ For this 
purpose it is necessary to consider the homogeneity of space, that is, 
the metric and the matter 
fields are independent of spatial coordinates, as a consequence one has
a finite number of degrees of freedom. Thus, the study of the associated 
quantum model becomes analogous to a supersymmetric quantum mechanical 
problem. The hope we have in these models is that they could give us the 
notion of the full quantum theory of SUGRA. The supersymmetric quantum 
cosmological models have been 
intensively studied with the hope to get a consistent quantum theory for the 
cosmological models. However, not all the results obtained in supersymmetric 
quantum cosmology have their counterpart in the full theory of 
SUGRA. Some of these problems have already been mentioned in two
extensive works.$^5$  

More recently, we have proposed a new approach to investigating the 
supersymmetric quantum cosmology.$^6$ In this approach we started 
with the action of the spatially hom\-ogeneous minisuperspace models and 
proceeded with
supersymmetrization. Because the starting action preserves the invariance 
under local time reparametrization, then the supersymmetric action must be
invariant under the extended local symmetry (supersymm\-etry). In order to 
have 
a local $n=2$ supersymmetry in $^6$, the odd ``time" parameter $\theta$ 
and its complex conjugate $\bar \theta$ were introduced. This involved 
introducing the superfield formulation, because superfields defined on 
superspace allow all the component fields in a supermultiplet to be 
manipulated simultaneously in a manner, that automatically preserves 
supersymm\- etry. This approach has the advantage of being simpler than the 
proposed models based on full SUGRA $^{4,5,7}$, and by means of 
this local symmetry procedure it gives the corresponding fermionic partners 
in a direct manner. Using the superfield procedure we have 
constructed the superfield action for all Bianchi-type models.$^8$
The inclusion
of the real scalar matter fields, as well as the parameter of spontaneous 
breaking of local supersymmetry were discussed in Ref. 9. Using the last 
results a normalizable wavefunction was obtained for the FRW model 
in Ref. 10. Although these models do not attempt to describe the real world, 
they keep many features occurring in four-dimensional space-time, 
which could really be studied in the quantum versions of simplified models.

The most physically interesting 
case is provided by $n=4$ local supersymmetry, since it can be applied 
to the description of the systems resulting from the ``realistic'' 
$N=1$, $D=4$ SUGRA subject to an appropriate dimensional reduction 
down to $D=1$.  

In this work we extend the transformations of time reparametrization to the
$n=4$ local supermsymetry with  $SU{(2)}_{local}\otimes SU{(2)}_{global}$ 
internal symmetry for all Bianchi-type-A cosmological models, and we give a 
procedure for including other matter fields in a sistematic way. This paper generalizes the $n=4$ construction described in Ref. 11. The extension presented
 is desirable for two reasons: 1) supersymmetric minisuperspace
models are related to full $N=1$, $D=4$ dimensional SUGRA by dimensional
reduction to $(1+0)$ dimensions.$^8$ By such reduction $N=1$, $D=4$ SUGRA
goes over to an $n=4$, $D=1$ supersymmetric model; 2) the gravitational 
field should be coupled to a supersymmetric matter field, like a complex 
scalar field.

\section{ $n=4$ superconformal transformations and the action}

The Bianchi models are the most general homogeneous cosmologies with a
3-dimensional group of isometries. These groups are in a one-to-one
correspondence with 3-dimensional Lie algebras, which were classified long
time ago by Bianchi. There are nine distinct 3-dimensional Lie algebras, and 
consequently nine types of Bianchi cosmologies. The 3-metric for each of these
models can be written in the generalized coordinates 

\begin{equation}
ds^2 = G_{\mu \nu}(q^{\lambda}) dq^\mu dq^\nu,
\label{1}
\end{equation}
where the generalized coordinates 
$ q^{\lambda}(\alpha, \beta_{+}, \beta_{-})$ with $\nu=0,1,2$ span the
minisuperspace with the metric $G_{\mu\nu}$, which
we may choose as flat, making use of the fact that the metric in 
minisuperspace is fixed only up to an arbitrary conformal
factor, written as $\exp[2 \omega(q)]$. All Bianchi-types models are
conformally flat, i.e. its metric takes the form

\begin{equation}
G_{\mu \nu}(q) = e^{2 \omega(q)} G^{(0)}_{\mu \nu},
\label{2}  
\end{equation}
with $G^{(0)}_{\mu \nu} = diag(-1, 1, 1)$. The inverse of this conformal factor
appears in the potential of each Bianchi model,

\begin{equation}
U(q) = e^{-2 \omega (q)} U_0(q),
\label{3}
\end{equation}
with the potential $U_0 = -^{(3)}g^{(3)} R$, where $^{(3)}g$ is the 
determinant and $^{(3)}R$ is the scalar curvature of the 3-metric. The action
for the Bianchi type models may be written as

\begin{equation}
S = \frac{1}{2} \int \left\{ \frac{1}{N} G_{\mu \nu}(q) {\dot q}^{\mu} 
{\dot q}^{\nu} + NU(q)\right\} dt.
\label{4}
\end{equation}
The lapse function $N(t)$ and the coordinates $q^{\mu}(t)$ depend on the
time parameter $t$ only. The action (\ref{4}) is invariant under 
reparametrization of $t^{\prime} \to t + a(t)$, if the transformations of 
$q^{\mu}$ and $N(t)$ are defined as 

\begin{eqnarray}
\delta q^{\mu}(t) &=& a(t) {\dot q}^{\mu}, \qquad \delta N(t) = (aN)^{.}.
\label{5}
\end{eqnarray}
That is, $q^{\mu}(t)$ transforms as a scalar and $N(t)$ as a one-dimensional
vector and its dimensionality is inverse to that of $a(t)$. It is easy
to see, that the action (4) gives a simple one-dimensional model for the 
somehow interacting homogeneous ``matter" field $q^{\mu}$ and gravity field 
$N(t)$. Using the superfield formalism the $n=2$ local SQM for cosmological 
models was constructed in Ref. 6,8.

In the action (\ref{4}), $U(q)$ corresponds to the potential of each 
Bianchi-type models. This potential may be written as

\begin{equation}
U(q) = \frac{1}{2} G^{\mu\nu}\frac{\partial \phi}{\partial q^{\mu}}
\frac{\partial \phi}{\partial q^{\nu}} = \frac{1}{2}G_{\mu\nu}m^{\mu}m^{\nu}.
\label{6}
\end{equation}
Thanks to this relation, the hidden symmetry of the cosmological models was 
found (see Ref. 7). This allows to construct a corresponding SQM. However, in 
this case the supersymmetry is global. It is natural to demand, 
that any cosmological action is invariant under local transformations. For this
reason the more extended symmetry must be local.

To construct  the 
superfield action in the world-line superspace $(t,\theta^a, \bar\theta_a)$ 
[with $t$ being a time parameter, and $\theta^a$ and 
$\bar\theta_a=(\theta^a)^{\ast}$, where $a=1,2 $ is  an $SU(2)$ index, being 
two complex Grassmann coordinates] one introduces a real ``matter'' superfield
 ${\bf Q}^{\mu}(t,\theta^a, \bar\theta_a)$ and a
 world-line supereinbein ${I\!\!N}(t,\theta^a, \bar\theta_a)$ which has
 the following properties with respect to the $SU(2)$ $n=4$
 superconformal transformations of the world-line superspace $^{12}$
 \footnote{Our conventions for spinors are as follows:
 ${\theta}_{a} ={\theta}^{b}{\varepsilon}_{ba},\; {\theta}^{a}=
 {\varepsilon}^{ab}{\theta}_{b},\; {\bar \theta}_{a}={\bar
 \theta}^{b}{\varepsilon}_{ba},\; {\bar \theta}^{a}=
 {\varepsilon}^{ab}{\bar \theta}_{b},\; {\bar \theta}_a =
 (\theta^a)^\ast,\; {\bar \theta}^a = -(\theta_a)^\ast,\; (\theta
 \theta)\equiv \theta^a \theta_a = -2 \theta^1 \theta^2,\;
  (\bar \theta \bar \theta)\equiv \bar \theta_a \bar \theta^a = (\theta
 \theta)^\ast,\;  (\bar \theta \theta) \equiv \bar \theta_a \theta^a,\;
  \varepsilon^{12}=- \varepsilon^{21}= 1,\;
 \varepsilon_{12} = 1$.}

 \begin{eqnarray}
 \delta t &=& \Lambda-\frac{1}{2}\theta^aD_a\Lambda - \frac{1}{2}
 \overline{\theta}_a\overline{D}^a\Lambda,\qquad
 \delta\theta^a = i\overline{D}^a\Lambda,\qquad
 \delta\overline{\theta}_a=iD_a\Lambda,
 \label{6}
 \end{eqnarray}
 \begin{equation}
 \delta {\bf Q}^{\mu} = -\Lambda\dot {\bf Q}^{\mu} + \dot\Lambda{\bf Q}^{\mu} -
i(D_a\Lambda)
 (\overline{D}^a{\bf Q}^{\mu})-i(\overline{D}^a\Lambda)( D_a{\bf Q}^{\mu}),
 \label{7}
 \end{equation}
 \begin{equation}
 \delta {I\!\!N} = -\Lambda\dot{{I\!\!N}}-\dot\Lambda {I\!\!N}-
 i(D_a\Lambda)(\overline{D}^a {I\!\!N})-i(\overline{D}^a\Lambda)
 (D_a {I\!\!N})
 \label{8},
 \end{equation}
 where the overdot denotes the time derivative $d/dt$. The
 transformation law (\ref{7}) for the superfield ${\bf Q}^{\mu}$ shows that 
this superfield is a vector superfield in the one-dimensional $n=4$ superspace,
while the superfield ${I\!\!N}{\bf Q}^{\mu}$ is a scalar.$^{12}$

The superfield ${\bf Q}^{\mu}$ obeys the quadratic constraints 
 \begin{eqnarray}
 \lbrack D_a , \overline {D}^a\rbrack {\bf Q}^{\mu} &=& -4m^{\mu}, \qquad\qquad
 D^a D_a {\bf Q}^{\mu} = 0, \qquad\qquad
 \overline{D}_a \overline{D}^a {\bf Q}^{\mu} = 0,
 \label{9}
 \end{eqnarray}
and it is irreducible representation of $n=4$ supersymmetry.$^{12}$
The vector $m^{\mu}$ depends on the concrete 
cosmological model in consideration. For example, in the case
of the Friedmann-Robertson-Walker model $m^\mu$ has the form 
$m = {\sqrt k}/2 $, where $k$ takes the value $1,0,-1$, and the metric 
$G_{\mu\nu}$ has one component $G_{00} = - R$ (see Ref. 11).
 
The superfield ${I\!\!N}$ obeys the constraints
 \begin{eqnarray}
 \lbrack D_a , \overline{D}^a\rbrack \frac{1}{{I\!\!N}} &=& 0, \qquad
 D^a D_a \frac{1}{{I\!\!N}} = 0,
 \qquad \overline {D}_a \overline {D}^a \frac{1}{{I\!\!N}} =0,
 \label{10}
 \end{eqnarray}
which are imposed in order to have a one-to-one correspondence between
the number of transformation parameters and that of fields,  
and\\
\begin{center}
 $
 D_a = \frac{\partial}{\partial \theta^a}- \frac{i}{2} \overline{\theta}_a
 \frac{\partial}{\partial t} ,
 \quad\quad\quad
 \overline{D}^a = \frac{\partial}{\partial \overline{\theta}_a}
 - \frac{i}{2} \theta^a \frac{\partial}{\partial t},
 $\\
\end{center}
 are the supercovariant derivatives. The infinitesimal superfield
$\Lambda$, which appears in (7-9)
 \begin{eqnarray}
 \Lambda(t,\theta,\overline{\theta})&=& a(t)
 +\theta^a \overline{\alpha}_a(t)
 -\overline{\theta}_a \alpha^a(t)
 + \theta^a(\sigma^i)_a^b \overline{ \theta}_b b_i(t)
 \nonumber\\
 &&+\frac{i}{4}(\theta \theta) \overline{\theta}_a 
\dot{\overline \alpha}^a(t) -
 \frac{i}{4}(\overline{\theta} \overline{\theta}) \theta^a {\dot \alpha}_a(t)
 +\frac{1}{16}(\theta\theta)(\overline{\theta} \overline{\theta}) \ddot{a}(t),
 \label{12}
 \end{eqnarray}
 contains the parameters of the local time reparametrizations $a(t)$,
 local supertranslations $\alpha(t)$, $\overline \alpha(t)$, $b_i(t)$ 
 being a local SU(2) parameter of the world-line superspace. 
 
The constraints (10) can be explicitly solved,
 the solution is described by the superfield
 \begin{eqnarray}
 {\bf Q}^{\mu}(t,\theta,\overline{\theta})&=& q^{\prime \mu}(t)
 +\theta^a {\overline \lambda}_a^{\prime \mu}(t)
 -{\overline \theta}_a \lambda'~^{a \mu}(t) +
 \theta^a(\sigma_i)_a~^b {\overline \theta}_b F^{i \prime \mu}(t)
 + m^{\mu}(\theta \overline{\theta}) \label{13}\\
 &&+\frac{i}{4}(\theta \theta){\overline \theta}_a
{\dot{\overline \lambda}}^{\prime a \mu}
 - \frac{i}{4}(\overline{\theta} \overline{\theta}) 
\theta^a {\dot \lambda}^{\prime \mu}_a
 +\frac{1}{16}(\theta\theta)(\overline{\theta} \overline{\theta}) 
{\ddot q}^{\prime \mu}(t).\nonumber 
\end{eqnarray}
This superfield contains one bosonic field $q^{\mu}$ and 
the Grassmann-odd fermionic fields (they are four). $\lambda^a(t)$ and 
${\overline \lambda}_a(t)$ are their superpartner spin degrees of freedom, 
and  $F_a^b = {(\sigma^i)_a^b} F_i$ are three auxiliary fields,
where $(\sigma^i)_a^b$ $(i=1,2,3)$ are the ordinary Pauli matrices.

 The constraints (11) are described by the superfield
 \begin{eqnarray}
 \frac{1}{{I\!\!N}}(t,\theta,\overline{\theta})&=&\frac{1}{N(t)}
 +\theta^a \overline{\psi}^{\prime}_a(t)
 -\overline{\theta}_a \psi^{\prime}~^a(t)
 +\theta^a (\sigma^i)_a~^b {\overline \theta}_b
 V'_i(t) \label{14}\\
 && +\frac{i}{4} (\theta \theta)\overline{\theta}_a 
\dot{\overline \psi}^{\prime}~^a(t)
 - \frac{i}{4} ({\overline\theta} {\overline \theta})\theta^a 
\dot{\psi}^{\prime}_a + \frac{1}{16}(\theta\theta)
({\overline \theta} {\overline \theta})
 \frac{d^2}{dt^2}\frac{1}{N(t)}.\nonumber
 \end{eqnarray}
 The superfield ${I\!\!N}$ describes an $n=4$
 world-line supergravity multiplet consisting of the einbein
 ``graviton" $N(t)$, two complex ``gravitinos" $\psi^{\prime a}(t)$
 and ${\overline\psi}^{\prime}_a(t)$, and the $SU(2)$ gauge field
 $V^{\prime}_i(t)$. The components of ${I\!\!N}$ play the role of Lagrange 
multipliers. Their presence mean that the dynamics of the model is subject 
to constraints.

The $n=4$ superfield action for the Bianchi-type cosmological models invariant 
under $n=4$ superconformal symmetry has the form $^{11,12}$

\begin{equation}
S= \frac{-8}{\kappa^2} \int dt d^2\theta d^2 \overline{\theta}
{I\!\!N}^{-1} A({I\!\!N} {\bf Q}^{\mu}),
\label{15}
\end{equation}
where $\kappa^2 = 8\pi G_N$, $G_N$ is the Newtonian constant of gravity.
The action (\ref{15}) is the most general superfield action, which can be 
constructed with 
respect to the $n=4$ conformal supersymmetry. $A({I\!\!N}{\bf Q}^{\mu})$ is an 
arbitrary function of the superfields ${I\!\!N}{\bf Q}^{\mu}$ called 
super\-potential. Note, that in the case of $n=4$ local supersymmetry it is 
sufficient to construct one invariant action possessing a minimal number of 
time derivatives, unlike of two invariants, a kinetic part and the potential 
one, as in the case of $n=2$ local supersymmetry.$^6$

So, integrating (\ref{15}) over the Grassmann coordinates $\theta$,
$\overline\theta$ and making the following redefinition of the
component fields
 
\begin{eqnarray}
\psi &=& N^{3/2}\psi^{\prime}, \quad
V_i = 2N(V^{\prime}_i + N (\psi^{\prime}\sigma_i \overline{\psi}^{\prime})),
\quad
\lambda^{\mu} = {\sqrt N}(\lambda^{\prime \mu} - q^{\mu} 
\psi^{\prime}), \label{16}\\
F^{\mu}_i &=& 2{\sqrt N}\lbrace F^{\prime \mu}_i - q^{\mu} V^{\prime}_i +
\frac{\sqrt N}{2}(\psi^{\prime} \sigma_i \overline \lambda^{\mu}) + 
\frac{\sqrt N}{2}(\lambda^{\mu}\sigma_i {\overline \psi}^{\prime}),\rbrace
\qquad
q^{\mu} = Nq^{\prime \mu}, \nonumber
\end{eqnarray}
one obtains the component action

\begin{eqnarray}
S &=& \int \left\{\frac{1}{2N} G_{\mu\nu} Dq^{\mu} Dq^{\nu} + 
iG_{\mu\nu}({\overline \lambda}^{\mu}{\tilde D}\lambda^{\nu} + \lambda^{\mu} 
{\tilde D}{\overline \lambda}^{\nu}) + \frac{1}{2} G_{\mu\nu} 
F_i^{\mu}F^{i\nu} - 2\sqrt {N} \Gamma_{\mu\nu\rho} 
\lambda^{\nu}(\sigma_i){\overline \lambda}^{\mu} F^{i\rho} \right.
\nonumber\\ 
&&\left. - 2 \Gamma{\mu\nu\rho}(\psi {\overline \lambda}^{\mu} 
{\overline \lambda}^{\nu} {\lambda}^{\rho} + 
{\overline \psi} \lambda^{\mu} {\lambda}^{\nu} 
{\overline \lambda}^{\rho}) - N(\partial_{\mu} \Gamma_{\nu\rho\sigma})
({\overline \lambda}^{\mu} {\overline \lambda}^{\nu} {\lambda}^{\rho}
{\lambda}^{\sigma}) - 2N G_{\mu\nu} m^{\mu} m^{\nu} \right. \label{17}\\ 
&& \left. - 4N \Gamma_{\mu\nu\rho} {\overline \lambda}^{\mu} 
{\lambda}^{\nu} m^{\rho} + 2G_{\mu\nu} m^{\mu}({\overline \lambda}^{\nu} 
\psi + {\overline \psi} {\lambda}^{\nu}) \right \}dt, \nonumber
\end{eqnarray}
where $Dq^{\mu} = {\dot q}^{\mu} - i({\overline \psi} \lambda^{\mu} +
{\overline \lambda}^{\mu} \psi)$ is the supercovariant derivative, 
${\tilde D} \lambda^{\mu} = D\lambda^{\mu} + \Gamma^{\mu}_{\nu \rho}
{\dot q}^{\rho} \lambda^{\nu}$, where $D\lambda^{\mu} = {\dot \lambda}^{\mu} - 
\frac{1}{2}V \lambda^{\mu}$ is the $SU(2)$ covariant derivative.
In order to give a geometrical form we have introduced in the action (\ref{17})
the special metric

\begin{eqnarray}
G_{\mu \nu}(q) &=& \frac{\partial^2 A}{ {\partial q^{\mu}} {\partial q^{\nu}}} 
\qquad
A(q^\mu) = A({I\!\!N} {\bf Q}^\mu){}|_{\theta, \overline \theta =0},
\label{18}
\end{eqnarray}
in this case the Christoffel connection takes the form

\begin{equation}
\Gamma_{\mu \nu \rho}(q) = \frac{1}{2} 
\frac{\partial^3 A(q)}{{\partial q^{\mu}}{\partial q^{\nu}}
{\partial q^{\rho}}}, \label{19}
\end{equation}
and the Riemann curvature tensor 
\[
R_{\mu \nu, \rho \sigma} = \Gamma_{\mu \sigma}^{\eta} 
\Gamma_{\eta \nu \rho} - \Gamma_{\mu \rho}^{\eta} \Gamma_{\eta \nu \sigma}.
\]

In the action (\ref{17}) the components $F_i$ of the superfield ${\bf Q^{\mu}}$
appear without derivatives and, therefore, they are non-dynamical variables. 
We can eliminate $F_i$ by means of their equation of motion. Solving the
equation of motion of the auxiliary fields $F_i$ and substituting the
solution back into Eq.(\ref{17}) we obtain the component action. From the
component action we derive the first-class constraints varying it with
respect to $N(t), \psi(t), {\overline \psi(t)} $ and $V_i(t)$, respectively

\begin{eqnarray}
H_0 &=& \frac{\kappa^2}{2} G^{\mu \nu} P_{\mu} P_{\nu} + 
2G_{\mu \nu} m^{\mu} m^{\nu} 
+ 4 D_{\mu}m^{\nu} {\overline \lambda}^{\mu} \lambda_{\nu}  
\nonumber\\
&& - R_{\mu \nu,\rho \sigma}{\overline \lambda}^{\mu}
{\overline \lambda}^{\sigma}
\lambda^{\nu} \lambda^{\rho} - R_{\sigma \rho,\nu \mu} 
{\overline \lambda}^{\mu} \lambda^{\sigma}
{\overline \lambda}^{\nu} \lambda^{\rho} \label{21}\\
&& - D_{\mu} \Gamma_{\nu \rho \sigma}({\overline \lambda}^{\mu} 
{\overline \lambda}^{\nu})(\lambda^{\rho} \lambda^{\sigma}), \nonumber
\end{eqnarray}

\begin{equation}
{\overline{Q}}_a = {\overline \lambda}_a^{\mu} P_{\mu} - 2i G_{\mu \nu} 
{\overline \lambda}_a^{\mu} m^{\nu} + i\Gamma_{\mu \nu \rho} 
{\overline \lambda}_a^{\mu}
{\overline \lambda}^{\nu} \lambda^{\rho},
\label{23}
\end{equation} 

\begin{equation}
{Q}^b = \lambda^{b \mu}P_{\mu} + 2iG_{\mu \nu} \lambda^{b \mu} m^{\nu} + 
i\Gamma_{\mu \nu \rho} {\overline \lambda}^{\mu b} \lambda^{\nu} 
\lambda^{\rho},
\label{24}
\end{equation}
 and
\begin{equation}
{\cal F}_i = G_{\mu \nu}\lambda^{\mu b} (\sigma_i)_b^a {\overline 
\lambda}_a^{\nu},
\label{25}
\end{equation}
where $H_0$ is the Hamiltonian of the system, $Q^a$ and ${\overline Q}_a$ are 
the supercharges, and ${\cal F}_i$ is the generator of $SU(2)$ rotations.

So, following the standard procedure of quantization of the system with 
bosonic and fermionic degrees of freedom, we introduce the canonical Poisson
brackets

\begin{eqnarray}
\lbrace q^{\mu}, P_{\nu} \rbrace &=& \delta^{\mu}_{\nu}, \qquad 
\lbrace \lambda^{a \mu}, \pi_{(\lambda)b \nu} \rbrace = 
- \delta^a_b \delta^{\mu}_{\nu}, 
\qquad
\lbrace {\overline \lambda}_a^{\mu}, \pi^b_{({\overline \lambda})\nu} 
\rbrace = -\delta_a^b \delta^{\mu}_{\nu},\label{26}
\end{eqnarray}
where $P_{\mu}, \pi_{(\lambda)a \mu}$ and 
$\pi^a_{({\overline \lambda}) {\mu}}$ 
are the momenta conjugated to $q^{\mu}, \lambda^{ \mu}$ and 
${\overline \lambda}^{\nu}$ respectively. From the explicit form of the momenta

\begin{equation}
P_{\mu} = \frac{1}{\kappa^2} G_{\mu \nu} \lbrace{\dot q}^{\nu} -
i\kappa({\overline \psi}_a \lambda^{a \nu} - 
{\overline \lambda}^{\nu}_a \psi^a) \rbrace 
\label{27}
\end{equation}

\begin{eqnarray}
\pi_{(\lambda)a \mu} &=& -iG_{\mu \nu}{\overline \lambda}^{\nu}_a, \qquad 
\pi^a_{(\overline \lambda)\mu}= -iG_{\mu \nu}\lambda^{a \nu},
\label{28}
\end{eqnarray}
one can conclude, that the system possesses the second-class fermionic
constraints

\begin{eqnarray}
\Pi_{(\lambda)a \mu} &=& \pi_{(\lambda)a \mu} + 
iG_{\mu \nu} {\overline \lambda}^{\nu}_a, 
\qquad
\Pi^b_{(\overline \lambda) \mu} = \pi^b_{(\overline \lambda)\mu} + 
iG_{\mu \nu}{\lambda}^{b \nu},
\label{29}
\end{eqnarray} 
since

\begin{equation}
\lbrace \Pi^a_{(\overline \lambda)\mu}, \Pi_{(\lambda)b \nu}\rbrace = 
-2iG_{\mu \nu}\delta^a_b.
\label{30}
\end{equation}
Therefore, the quantization has to be done using the Dirac brackets, defined
by any of two functions $F$ y $G$ as

\[
\lbrace F, G \rbrace^{\ast} = \lbrace F, G \rbrace - 
\lbrace F, \Pi_a \rbrace \frac{1}{\lbrace \Pi_a, \Pi_b \rbrace }
\lbrace \Pi_b, G \rbrace.
\]
As a result, we obtain the following Dirac brackets for the canonical
variables

\begin{eqnarray}
\lbrace q^{\mu}, P_{\nu} \rbrace^{\ast} &=& \delta^{\mu}_{\nu}, \qquad
\lbrace \lambda^{a \mu}, {\overline \lambda}^{\nu}_b \rbrace^{\ast}= 
- \frac{i}{2}\delta^a_b G^{\mu \nu}, \nonumber\\
\lbrace \lambda^{a \mu}, P_{\nu} \rbrace^{\ast}&=& - \lambda^{a \rho} 
\Gamma_{\nu \rho}^{\mu},
\quad \lbrace {\overline \lambda}^{\mu}_a, P_{\nu} \rbrace^{\ast} = 
- {\overline \lambda}^{\rho}_a \Gamma_{\nu \rho}^{\mu}, \label{32}\\
\lbrace P_{\mu}, P_{\nu} \rbrace^{\ast} &=& 2iR_{\mu \nu,\rho \sigma} 
{\overline \lambda}^{\rho}
\lambda^{\sigma}. \nonumber
\end{eqnarray}

The supercharges and the Hamiltonian form the following $n=4$ SUSY QM algebra
with respect to the introduced Dirac brackets 

\begin{eqnarray}
\lbrace {\overline Q}_a , Q^b \rbrace^{\ast} &=& -i \delta_a^b H_0,  \quad 
\lbrace {\cal F}_j , {\cal F}_k \rbrace^{\ast} = \epsilon_{jkl} {\cal F}_l,
\nonumber\\
\lbrace {\cal F}_i , {\overline Q}_a \rbrace^{\ast} &=& \frac{i}{2} 
{(\sigma_i)}^c_a {\overline Q}_c, \quad 
\lbrace {\cal F}_i ,  Q^a \rbrace^{\ast} = - \frac{i}{2} {(\sigma_i)}^a_cQ^c.
\label{33} 
\end{eqnarray}

On the quantum level we replace the Dirac brackets by (anti)commutators 
using the rule 
\[
i\lbrace , \rbrace^{\ast} = \lbrace , \rbrace.
\]
one obtains the non-zero commutation relations

\begin{eqnarray}
\lbrack q^{\mu}, P_{\nu} \rbrack &=& i\delta_{\nu}^{\mu}, \quad
\lbrace \lambda^{a \mu}, {\overline \lambda}^{\nu}_b \rbrace = 
\frac{1}{2} \delta^a_b G^{\mu \nu} \label{35}\\
\lbrack P_{\mu}, \lambda^a_{\nu} \rbrack &=& i\Gamma_{\mu \nu \rho} 
\lambda^{a \rho}, \quad
\lbrack P_{\mu}, {\overline \lambda}^a_{\nu} \rbrack = 
i\Gamma_{\mu \nu \rho}{\overline \lambda}^{a \rho}, \nonumber\\ 
\lbrack P_{\mu}, P_{\nu} \rbrack &=& -2R_{\mu \nu,\rho \sigma}
{\overline \lambda}^{\rho} \lambda^{\sigma} \nonumber.
\end{eqnarray}

We observe that $P_{\mu}$ has properties of covariant momenta when acting on 
fermionic var\-iables $\lambda^{a \mu}$ and ${\overline \lambda}_a^{\mu}$.
The superalgebra of the constraints generates the 
$SU(2)_{local}\otimes SU(2)_{global}$ $n=4$ superconformal transformations of
the components of the superfields ${\bf Q}^{\mu}$.  
In the quantum theory the first-class constraints (\ref{21}-\ref{25}) 
associated with the invariance of the action (15,17) become conditions 
on the wave function $\Psi$ of the Universe. Therefore, any physically 
allowed states must obey the quantum constraints

\begin{eqnarray}
H_{0}\Psi&=& 0, \qquad Q^a\Psi =0, \qquad \overline{Q}_a\Psi=0, \qquad
{\cal F}_i\Psi=0.
\label{36}
\end{eqnarray} 
The quantum generators $H_{0}, Q^a, {\overline{Q}}_a$ and ${\cal F}_i$ form 
a closed superalgebra of the $n=4$ supersymmetric quantum mechanics

\begin{eqnarray}
\lbrace {\overline{Q}}_a, Q^b \rbrace &=& H_{0}\delta_a^b, \qquad
\lbrack {\cal F}_i, {\cal F}_j \rbrack = i\epsilon_{ijk} {\cal F}_k, \qquad
\lbrack {\cal F}_i, {\overline{Q}}_a \rbrack = 
-\frac{1}{2} (\sigma_i)_a^b {\overline{Q}}_b, \label{37}\\
\lbrack {\cal F}_i, Q^a \rbrack &=& \frac{1}{2} (\sigma_i)_b^a Q^b.
\nonumber
\end{eqnarray}
In order to obtain the quantum expression for the Hamiltonian $H_0$ and for
the supercharges $Q^a$ and $\overline{Q}_a$ we may solve the operator ordering
ambiguity, for example following the works.$^{11,13}$

\section{Conclusions}
On the basis of the local $n=4$ supersymmetry the superfield action for the 
Bianchi-types cosmological models is formulated. It is shown, that the 
action (\ref{15})
has the form of the localized version of $n=4$ supersymmetric quantum 
mechanics. Due to the quantum supersymmetric algebra (\ref{37}), the 
Wheeler-DeWitt equation, which is of the second-order, can be replaced by
the four first-order supercharge operator equations constituting its 
sup\-ersymmetric ``square root". 

It would be very interesting to consider the
interaction with matter fields and analize the spontaneous breaking of 
$n=4$ local supersymmetry. We hope, that for those more general supersymmetric 
cosmological models than in Ref. 10, we can find a normalizable wavefunction. 
The details of this study will be given elsewhere.

\noindent {\bf Acknowledgments.}
We are grateful to S. Odintsov, J. Socorro, H. Rosu and I. Lyanzuridi for 
their interest in this work and useful comments. This research was supported 
in part by CONACyT under the grants 2845-54E and 2845E. One of us (J.J.R.) 
would like to thank CONACyT for support under grant 000683.

\vspace{1.cm}
 
1. C.W. Misner, {\it in Magic Without Magic}, edited by
J.R. Klauder (Freeman, San Francisco, 1972).\\

2. C. Teitelboim, Phys. Rev. Lett. {\bf 38}, 1106 (1977).\\

3. S. Deser, J.H. Kay and K.S. Stelle, Phys. Rev. {\bf D16}, 
2448 (1977); E.S. Fradkin and M.A. Vasiliev, Phys. Lett. {\bf B72}, 70
(1977); M. Pilati, Nucl. Phys. {\bf B132}, 138 (1978); 
T. Jacobson, Class. Quantum Grav. {\bf 5}, 923 (1988).\\

4. A. Mac\'{\i}as, O. Obreg\'on and M.P. Ryan Jr, Class. Quantum
Grav. {\bf 4}, 1477 (1987).\\

5. P.D. D'Eath, {\it Supersymmetric Quantum Cosmology}, 
(Cambridge: Cambridge University Press, 1996); 
P.V. Moniz, {\it Supersymmetric Quantum Cosmology}, Int. J. Mod. 
Phys. {\bf A11}, 4321-4382 (1996).\\

6. O. Obreg\'on, J.J. Rosales and V.I. Tkach, Phys. Rev.
{\bf D53}, 1750 (1996).\\

7. R. Graham, Phys. Rev. Letters {\bf 67}, 1381 (1991); 
E.E. Donets, M. N. Tentyukov and M. M. Tsulaia, Phys. Rev.{\bf D59}, 023515 
(1999).\\

8. V.I. Tkach, J.J. Rosales and O. Obreg\'on, 
Class. Quantum Grav. {\bf 13}, 2349 (1996); V. I. Tkach, J. J. Rosales and J. Socorro, Class. Quantum Grav. {\bf 16}, 797 (1999).\\

9. V.I. Tkach, O. Obreg\'on and J.J. Rosales, Class. Quantum Grav. 
{\bf 14}, 339 (1997).\\

10. O. Obreg\'on, J.J. Rosales, J. Socorro and V.I. Tkach,
Class. Quantum Grav. {\bf 16}, 2861 (1999).\\

11. A. Pashnev, J.J. Rosales, V.I. Tkach and M. Tsulaia, 
Phys. Rev.{\bf D64}, 087502, (2001).\\

12. E. A. Ivanov, S.O. Krivonos and A. I. Pashnev, Class. Quantum 
Grav. {\bf 8} 19 (1991); E.E. Donets, A. Pashnev, V.O. Rivelles, D. Sorokin 
and M. Tsulaia, Phys. Lett. {\bf B 484}, 337 (2000).\\

13. V. de Alfaro, S. Fubini, S. Furlan and M. Roncadelli, Nucl. 
Phys.{\bf B 296},402 (1998); E.E. Donets, A. Pashnev, J.J. Rosales and 
M.M. Tsulaia, Phys. Rev.{\bf D 61}, 043512-1, (2000).\\

\end{document}